\documentclass[preprint,flushrt,showpacs,preprintnumbers,amsmath,amssymb]{revtex4}
\usepackage{graphicx}
\usepackage{latexsym}
\usepackage{color}
\usepackage{dcolumn}
\usepackage{bm}
\usepackage{natbib}

\begin{document}

\title{THE ORIGIN OF PRIMORDIAL MAGNETIC FIELDS}

\author{Rafael S. de Souza}
\email{Rafael@astro.iag.usp.br}
\author{Reuven Opher}%
 \email{Opher@astro.iag.usp.br}
\affiliation{IAG, Universidade de S\~{a}o Paulo, Rua do Mat\~{a}o
1226, Cidade Universit\'{a}ria, CEP 05508-900, S\~{a}o Paulo, SP,
Brazil.}

\begin{abstract}

 Magnetic fields of  intensities  similar to those  in our galaxy are also
observed in high redshift galaxies, where a mean field dynamo would
not have had time to produce them. Therefore, a primordial origin is
indicated. It has been suggested that magnetic fields were  created
at various primordial eras: during inflation, the electroweak phase
transition, the quark-hadron phase transition (QHPT), during the
formation of the first objects, and during reionization.  We suggest
here that the large scale fields $\sim \mu$G, observed in galaxies
at both high and low redshifts by Faraday Rotation Measurements
(FRMs), have their origin in the electromagnetic fluctuations that
naturally occurred in the dense hot plasma that existed just after
the QHPT. We evolve the predicted fields   to the present time. The
size of the  region containing a coherent magnetic field increased
due to the fusion  of smaller regions. Magnetic
fields (MFs) $\sim 10 \mu$G over a comoving $\sim 1$ pc region are
predicted at redshift \emph{z} $\sim 10$. These fields are orders of
magnitude greater than those predicted in previous scenarios for
creating primordial magnetic fields. Line-of-sight average magnetic
fields (MFs) $\sim$ $10^{-2}$ $\mu$G, valid for FRMs,  are obtained
over a 1 Mpc comoving region at the redshift \emph{z} $\sim$ 10.
 In the collapse to a  galaxy (comoving size $\sim$ 30 kpc) at \emph{z} $\sim$ 10, the fields
  are amplified to $\sim 10 \mu$G. This indicates that the MFs created immediately after the
 QHPT ($10^{-4}$ s),
 predicted by the Fluctuation-Dissipation Theorem,
  could be the origin of
  the $\sim \mu$G fields observed  by FRMs in galaxies at both high and
 low redshifts. Our predicted MFs are shown to be consistent with present
observations. We discuss the possibility that the predicted MFs
could cause non-negligible deflections of ultra-high energy cosmic
rays and help create the observed isotropic distribution of their
incoming directions. We also discuss the importance of the volume
average magnetic field predicted by our model  in producing the
first stars and in reionizing the Universe.

\end{abstract}

\maketitle

\section{Introduction}

The origin of galactic and extragalactic magnetic fields is one of
the most challenging problems in modern astrophysics
\citep[e.g.,][]{ree87,kro94}. Magnetic fields on the order of $\sim
\mu$ G are detected  in galaxies as well as in clusters of galaxies.
It is generally assumed that the coherent large scale $\sim \mu$G
magnetic fields observed in disk galaxies are amplified and
maintained by an $\alpha-\omega$ dynamo, which
 continuously generates new fields by the combined action of
differential rotation $(\omega)$ and helical turbulence $(\alpha)$.
However, the dynamo mechanism needs  seed magnetic fields and
sufficient time in order to amplify them.

There have been many attempts to explain  the origin of seed fields.
One of the most popular  is that they are generated by the Biermann
mechanism \citep{bie50}. It has been suggested that this mechanism
acts in diverse astrophysical systems, such as  large scale
structure formation \citep{pee67,ree72,was78}, cosmological ionizing
fronts \citep{gne00}, and formation of supernova remnants of the
first stars \citep{mir98}. Outflows is an additional means of filling
 protogalaxies with magnetic fields. For example,
 in Section III. G, we discuss outflows of magnetic fields
 from extragalactic jets, as suggested in \citep{jaf92}.

 Another suggestion for the origin of seed fields is that they were created by different mechanisms in the very early
Universe, before galaxy formation took place. For example, such
fields may have been created during the quark-hadron phase
transition (QHPT),
 when the Universe was at a temperature $T_{QHPT} \cong
10^{12} K$ (Section III. A), during the electroweak phase transition (Section III. B), or in the Inflation era (Section III. C).

One major difficulty with  most scenarios for the creation of
 magnetic fields in the very early primordial
Universe ($\ll$ 1 sec), such as those discussed in \S 3.1-3.3, is
the small  coherence lengths of the fields at redshifts \emph{z}
$\lesssim$ 10. The coherence length is limited by the radius of the
horizon at the time of the creation of the magnetic field.
  When expanded to the present time,  the coherence length is too small to explain the existing observed
large coherent magnetic fields on the order of the size of galaxies.

In this paper, we suggest that the observed magnetic fields have
their origin in the electromagnetic fluctuations in the hot dense
plasmas of the very early Universe. This is a natural way to create
magnetic fields and circumvents the problem of small coherence
lengths. The Fluctuation-Dissipation-Theorem predicts very large
magnetic fields in the equilibrium plasma immediately after the
QHPT. We evolve these fields to a redshift $\emph{z} \sim 10$, when
galaxies were beginning to form and find them to be sufficiently
strong  to
 explain the magnetic field observations  in both high  and low redshift
galaxies.

We investigate the magnitude of the present magnetic fields in
galaxies and the intergalactic medium created by the plasma
fluctuations shortly after the quark-hadron-phase transition (QHPT),
when the plasma properties are well understood. The magnetic fields,
created by the plasma fluctuations before the QHPT, are poorly
understood and we leave their evaluation for a future investigation.

Using the Fluctuation-Dissipation-Theorem (FDT),
\citet{oph97a,oph97b,oph99} studied the magnetic fluctuations as a
function of frequency in the primordial nucleosynthesis era and
found that they were very large, in particular, at zero frequency.
This can be compared with the black body prediction which has a zero
amplitude magnetic fluctuation at zero frequency.

\citet{taj92} suggested that the large magnetic fluctuations
predicted by the FDT at an early epoch did not dissipate, but
continued to exist to the present epoch and  now  contribute to the
dominant magnetic field. This scenario is investigated in detail
here. Since the largest magnetic fluctuations in the plasma occurred
shortly after the QHPT, we begin our calculations at this epoch.

Primordial magnetic fields can effect the incoming directions of
ultra-high energy cosmic rays (UHECRs) above $ 3\times 10^{18}$ eV.
In the last section (Section VI) we discuss the possible importance of our
predicted primordial magnetic fields on UHECRs.

 We review the observations of astrophysical magnetic fields in Section II
  Previous suggestions for creating primordial magnetic fields are
given in Section III. The creation of magnetic fields in the
fluctuations of the hot dense primordial plasma is discussed in
Section IV. In Section V , we discuss our model, based on the
analysis in Section IV . Our conclusions
 as well as a discussion of our results are presented in Section VI.

\section{Observations of Cosmic Magnetic Fields}

 The magnetic fields in  our Galaxy have been studied by several
 methods. Measurements of the Zeeman effect in the 21 cm radio line in galactic HI
 regions reveal  magnetic fields  $\simeq 2-10
 \mu$G. Similar values for the magnetic fields in other galaxies have been obtained
 from Faraday rotation surveys.

 Observations of a large number of Abell clusters have provided information on magnetic
 fields in clusters of galaxies \citep{pol05,vog03,gov02}. The typical magnetic field strength in
 the cluster  is $\sim 1-10\mu$G, coherent over $10-100$
 kpc.

 High resolution FRMs of high \emph{z} quasars
 allow for the probing of magnetic fields in the past.  A magnetic field  of $\sim \mu$ G in a
 relatively young spiral galaxy at \emph{z} = 0.395 was measured  by FRMs
 from the radio emission of the quasar PKS 229-021, lying behind the
 galaxy, at \emph{z} = 1.038 \citep{kro90}. Magnetic fields $\sim  \mu$G are also observed in $Ly\alpha$ clouds at
redshifts  \emph{z} $\sim$ 2.5 \citep{kro94}.

 \section{Previous Suggestions for Creating Primordial Magnetic Fields}

 There have been various scenarios suggested for the source of
 primordial magnetic fields. In this section, we review some of
 the most important ones.

 \subsection{Magnetic Fields Created at the Quark-Hadron Phase
 Transition}

 In the magnetogenesis scenario at the quark-hadron phase transition (QHPT),  proposed by
 \citet{qua89},  an electric field was created
 behind the shock fronts due to the expanding bubbles of the phase transition. The baryon asymmetry, which was presumed
 to have already been present, resulted in a positive charge on the baryonic component
  and a negative charge on  the leptonic component of the primordial plasma, so that the  charge neutrality of the Universe was preserved.
 As a consequence of  the difference between
 the equations of state of the baryonic and leptonic fluids,  a
  strong pressure gradient was produced by the passage
 of the shock wave, giving rise to a radial electric field behind
 the shock front.  \citet{qua89} estimated  the strength
 of the electric field to be
 \begin{equation}
 eE \simeq
 15\left(\frac{\epsilon}{0.1}\right)\left(\frac{\delta}{0.1}\right)\left(\frac{kT_{QHPT}}{150MeV}\right)\left(\frac{100
 cm}{\textit{l}}\right)\frac{keV}{cm},
 \end{equation}
 where $\epsilon$ is the ratio of the energy density of
 the two fluids, $\delta \equiv (\textit{l}\Delta p/p)$, $\Delta$p is the pressure gradient, and \textit{l}
 is the average comoving distance between the nucleation sites.
  They suggested that non- negligible  fields were  produced when shock fronts collided, giving rise to turbulence
 and vorticity on scales of order \textit{l}. It was found that the magnetic field
 produced on the comoving scale $\sim 1$ AU  has a present magnitude
 $\sim$ $2\times10^{-17}$ G.

  \citet{che94} showed that
 strong magnetic fields might have been  produced during the coexistence
 phase of the QHPT, during which  a baryon
 excess builds up in front of the bubble wall as a
 consequence of the difference of the baryon masses in the quark
 and hadron phases. In this scenario, magnetic fields were generated
 by the peculiar motion of the dipoles, which arose from the
 convective transfer of the latent heat released by the expanding
 walls. The field created at the QHPT  was estimated by
 \citet{che94} to be  $\simeq 10^{-16}$ G at the present epoch, on a
comoving coherence length $\simeq 1 pc$.
  On a comoving galactic length scale, they estimated the field to be $\simeq 10^{-20}$ G.

 \citet{sig97} predicted a present magnetic field $\simeq10^{-9}$ G.
 However, they used
   very special conditions, such as efficient
 amplification by hydromagnetic instabilities
 during the QHPT.

 \subsection{Magnetic Fields From The Electroweak Phase Transition}

 There have been some suggestions made for the origin of
 primordial magnetic fields based on the electroweak phase
 transition (EWPT).
 A first order EWPT could possibly have generated  magnetic fields \citep{bay96,sig97}. During the
 EWPT, the gauge symmetry broke down from the
 electroweak group $SU(2)_{L}\times U(1)_{Y}$ to the
 electromagnetic group $U(1)_{EM}$. The transition appears to have been
 weakly first order, or possibly second order, depending upon various  parameters, such
 as the mass of the Higgs particle \citep{bay96,boy01}. If it were
 first order, the plasma would have supercooled below the electroweak
 temperature, $\simeq 100$ GeV. Bubbles of broken symmetry would
 have nucleated and expanded, eventually filling the Universe. At the time of the
 EWPT, the typical comoving size of the Hubble radius
   and the temperature were $ L_{H} \approx$ 10 cm and  $ T_{H} \approx$ 100 GeV, respectively. A
comoving  bubble of size $L_{B} = f_{B}L_{H}$ would have been
created with $f_{B}\simeq
 10^{-3}-10^{-2}$ \citep{bay96}.
  The fluids would have become turbulent when two walls collided. Fully developed
 MHD turbulence would have led rapidly to equipartition of the field energy up
 to the scale of the largest eddies in the fluid, assumed to have
 been
 comparable to $L_{B}$. The magnetic field strength at the EWPT
 would have been
 \begin{equation}
 B\simeq (4\pi\epsilon)^{1/2}(T_{EW})T^{2}_{EW}\left (\frac{v_{wall}}{c} \right)^{2}
 \simeq(7 \times 10^{21}-2\times 10^{24}) G,
 \end{equation}
 where $\epsilon = g_{*}aT^{4}_{EW}/2\simeq4\times10^{11} GeV
 fm^{-3}$ is the energy density at the time of the EWPT \citep{wid02}.

 Magnetic fields could  also have arisen in cosmological phase transitions even
 if they were of second order \citep{vac91}. In the standard model, the EWPT
 occurred when the Higgs field $\phi$ acquired a vacuum expectation value
$\eta$. To estimate the
 field strength on larger scales, \citet{vac91} assumed that $\phi$
 executed a random walk on the vacuum manifold with step size
 $\xi$. Over a distance $L = N\xi$, where N is a large number, the
 field $\phi$ changes on the average by $N^{1/2}\eta^{-1}$. On a
 comoving
 galactic scale, L = 100 kpc, at
 the recombination era  ($\emph{z} \sim 1100$), \citet{vac91} found a magnetic field $\simeq 10^{-23}$ G.
\subsection{Magnetic Fields Generated During Inflation}

Inflation naturally produced effects on large scales, very much
larger than the Hubble horizon, due to microphysical processes
operating in a causally connected volume before inflation
\citep{tur88}. If electromagnetic quantum fluctuations were
amplified during inflation, they could appear today as large-scale
coherent magnetic fields. The main obstacle to  the inflationary
scenario is the fact that in a conformally flat metric, such as the
Robertson Walker,  the background gravitational field does not
produce relativistic particles if the underlying theory is
conformally invariant \citep{par68}. This is the case for photons,
for example, since classical electrodynamics is conformally
invariant in the limit of vanishing fermion masses (i.e., masses
much smaller than the inflation energy scale). Several ways of
breaking conformal invariance have been  proposed. \citet{tur88}
considered three possibilities:
\begin{enumerate}
\item introducing a gravitational coupling, such as  $RA_{\mu}A^{\mu}$ or
$R_{\mu \nu}A^{\mu}A^{\mu}$, where R is the Ricci scalar, $R_{\mu
\nu}$ the Ricci tensor, and $A^{\mu}$ is the electromagnetic field.
These terms break gauge invariance and give the photons an effective
time-dependent mass. \citet{tur88} showed that for some suitable
(though theoretically unmotivated) choice of parameters, such a
mechanism could give rise to galactic magnetic fields, even without
invoking the galactic dynamo; \item introducing terms of the form
$R_{\mu \nu \lambda \kappa}F^{\mu\nu}F^{\lambda\kappa}/m^{2}$ or
$RF^{\mu\nu}F_{\mu\nu}$, where m is some mass scale, required by
dimensional considerations. Such terms arise due to one loop vacuum
polarization effects in curved space-time. They can account,
however, for only  a very small primordial magnetic field; and
\item  coupling of the photon to a charged
field that is not conformally coupled or  anomalous coupling to a
pseudoscalar field.
\end{enumerate}

\citet{dol93} proposed a model invoking a spontaneous breaking of
the gauge symmetry of electromagnetism, implying non-conservation of
the electric charge in the early evolution of the Universe.

\subsection{Generation of the Primordial Magnetic Fields During The Reionization Epoch}

\citet{gne00} investigated the generation of magnetic fields by the
Biermann battery in cosmological ionization fronts, using
simulations of  reionization by stars in protogalaxies. They
considered two mechanisms: 1) the breakout of ionization fronts from
protogalaxies; and 2) the propagation of ionization fronts through
high-density neutral filaments. The first mechanism was dominant
prior to the overlapping of ionized regions $(\emph{z} \approx 7)$,
whereas the second mechanism continued to operate  after that epoch
as well. After overlap, the  magnetic field strength at
$\emph{z}\approx5$ closely traced the gas density and  was highly
ordered on comoving megaparsec scales. The present mean  field
strength was found to be $\approx10^{-19}$ G in their simulation.
Their results corroborate those of \citet{sub94}.

\subsection{Generation of Magnetic Fields Due to Nonminimal
Gravitational-Electromagnetic Coupling After Recombination}

 The generation of magnetic fields by nonminimal coupling was investigated by \citet{oew97}. From General
 Relativity, it can be shown that if we have a mass spinning at the
 origin, the space time metric $g_{oi}$ is equal to the vector
 product of the angular momentum L and the radial vector r,
 times $2G/c^{3}r^{3}$, where G is the gravitational constant.
 \citet{oew97} suggested that the  magnetic field created is
 proportional to the curl of $g_{0i}$, where the proportionality
 constant  $\sim(G)^{1/2}/2c$ was used, based on the data of the
 planets in our solar system \citep{sch80,bla47}.

 Angular momentum in galaxies has been previously  suggested to have been created by  tidal
 torques between  protogalaxies \citep{pee69,efs79,whi84,bar87}. The spin parameter
 $\lambda$ is defined as the ratio of the angular velocity of
 the protogalaxy to the angular velocity required for the
 protogalaxy to be supported  by rotation alone. Numerical
 simulations find $\lambda \sim 0.05$, while observations of
 spiral galaxies show $\lambda \sim 0.5$. Since $\lambda$ is
 proportional to the square root of the binding energy, it increases by a factor of ten in the formation of a galaxy
 due to an increase  of the binding energy by a factor of 100 (i.e., the
 radius of the protogalaxy decreases by a factor of 100).

 In their calculations, \citet{oew97} investigated models in
 which the angular momentum of a galaxy increased until the decoupling
 redshift $\emph{z}_{d}$ and remained constant thereafter.  At the decoupling
 redshift, the spin parameter was
  $\lambda \sim 0.05$.  They
 found present galactic magnetic fields $\sim$ 0.58 $\mu$G for a decoupling redshift $\emph{z}_{d}$ =
 100 and noted that galactic magnetic fields $\sim \mu$G could be produced
by
 this mechanism
 without the need for dynamo amplification.

\subsection{Creation of Magnetic Fields From Primordial Supernova
Explosions}

 Primordial supernova explosions could also be  the origin of magnetic fields in the
 Universe \citep{mir96,mir97,mir98}. The scenario investigated was a generic multicycle
 explosive model, in which a  Population III object collapsed and then
 exploded, creating a shock. Matter was swept up by the shock, increasing the
 density  by a factor of 4 (for the case of a strong shock). This matter was heated
 to a high temperature, which  then cooled
 down. Eventually spheres of radii of approximately  half the shell
thickness
 formed and subsequently collapsed into  Population III stars. They then  exploded, starting a new cycle.
 The supernova shells produced eventually
 coalesced. It was assumed that the  gradients of
 temperature and density in the resultant shell were not  parallel and that, therefore, a magnetic field
 was created due to the Biermann mechanism. The rate of change of the
 magnetic field with time is equal to the vector product of the
 density gradient and the temperature gradient times $4\pi
 k_{B}/\pi en$, where n is the particle density and $k_{B}$  is the
 Boltzmann
 constant.  It was found that this process creates a
 galactic seed magnetic field $\sim 10^{-16}$G, which could be later
 amplified by a dynamo mechanism.

 \subsection{The Origin of Intergalactic Magnetic Fields Due to Extragalactic Jets}

 \citet{jaf92} suggested that the large-scale magnetization of the
intergalactic medium is due to electric current carrying
extragalactic jets, generated by active galactic nuclei at high
\emph{z}. The action of the Lorentz force  on the return current
expanded it into the intergalactic medium. Magnetic fields created
by these
 currents were identified as  the origin  of the
intergalactic magnetization. They found magnetic fields $\sim$
$10^{-8}$G over comoving Mpc regions.

\subsection{Magnetic Field Generation from Cosmological
Perturbations}

Another class of magnetic field generation studies are those based
on cosmological perturbations. A recent article on this subject is
that of \citet{tak05}. They studied the evolution of a three
component plasma (electron, proton, and photon), taking into account
cosmological perturbations. The
 collision term between electrons and photons was evaluated up to
 second order and was shown to induce a magnetic field $\sim 10^{-19}$ G on a 10 Mpc comoving scale at
decoupling.

\subsection{Magnetic Field Generation Due to Primordial Turbulence}

Turbulence has been suggested as the primordial source of magnetic
fields. \citet{ban04} has made a detailed study of this scenario. We
summarize here their analysis and results and compare them with the
analysis and results of the present paper.

It was assumed by \citet{ban04} that non-standard out-of-equilibrium
stochastic magnetic fields were created at high cosmic temperatures
T $\sim 100$ MeV - 100 GeV, corresponding to quark-hadron or
electroweak phase transitions. Their numerical simulations were
performed using the ZEUS-3D code. Gaussian random fields were used
to create the non-standard initial turbulent fluctuations. A
power-law with distance $l$ was assumed for the magnetic amplitudes,
($B \propto l^{-n}$, n = 1-2). The initial stochastic velocity field
was generated in the same way as the initial magnetic field. A
correlation length scale L was defined which contains most of the
magnetic and fluid kinetic energy. The dissipation of the energy
into heat occurs via energy cascading from large eddies ($\sim$ L)
to small eddies ($\sim l_{diss}$).

Ever since the work of Kolmogorov, it has been known that cascading
of energy occurs due to eddies on a scale $l$ breaking up into
smaller eddies ( $\sim l/2$). Typical energy dissipation times due
to the eddy flows from large to small flows are given by the eddy
turnover time on the scale L. In the article of \citet{ban04}, the
turnover time on the scale L is comparable to the Hubble time. Thus
the  turbulent energy introduced in the magnetogenesis era is
dissipated in one Hubble time. For example, the dissipation time is
$\sim 10^{-4}$ s for the quark-hadron transition.

The predicted present magnetic field in this turbulent eddy scenario
depends on the turbulent spectrum assumed at the quark-hadron or
electroweak phase transitions.  \citet{ban04} found the present
magnetic field  to be correlated with the comoving correlation
length $L_{c}$:
 B $\simeq 5 \times 10^{-15} L_{c}$ G, where $L_{c}$ is measured in pc.
  Typically, it was found that $L_{c} \sim 10^{-2}$ (Eq. (52) in \citep{ban04}).
  Thus, the turbulent eddy scenario, with the large eddy energy transfers to
  small scales, where energy dissipation rapidly occurs, typically
  predicts
  $\sim 10^{-16}$G on $\sim 10^{-2}$ pc scales. There can be
  substantial energy transfer   to larger scales if the turbulent
  magnetic field possesses some magnetic helicity \citep{bra96}.

The above can be compared with the magnetogenesis in  the present
paper,  due to the
 natural fluctuations in thermal equilibrium plasmas. Initially, the
   magnetic fluctuations had an average
  size $\bar{\lambda} = (7\pi/3 (c/\omega_{p})$ [Eq. (15)], where
  $\omega_{p}$ is the plasma frequency and c is the velocity of light.
  They have an average intensity $\langle\bar{B}^{2}\rangle/8\pi$ =
  $(T/2)(4\pi/3)/\bar{\lambda}^{-3}$, where T is the temperature.
  Describing  the magnetic fluctuations  as  dipoles,
  the magnetic field over a distance $l$ due to the randomly oriented
  magnetic fields follows a power law: $B =
  \bar{B}(\bar{\lambda}/l)^{3/2}$. This power law dependence is
  similar to the power law dependence in the  turbulent
  magnetogenesis model, but without a   transfer of energy  from large to small
  scales.  For a  thermal
  equilibrium plasma, the eddy turnover velocity of size $l$
  is  the thermal rotation velocity of the mass of plasma with a
  diameter $l$. In the power-law spectrum $B=
  \bar{B}(\bar{\lambda}/l)^{3/2}$ with $l > \bar{\lambda}$, the eddy
  turnover time is greater than the Hubble time. There is thus
  negligible energy transfer from the large scale $l$ to the small
  scale $\bar{\lambda}$. On the small scale $\bar{\lambda}$,
  dissipation,
  has already been taken into account by the
  Fluctuation-Dissipation Theorem.

  The present predicted magnetic field in our magnetogenesis model can be compared with the
  predicted magnetic field of the turbulent eddy model. Whereas, in
  the turbulent eddy model, a present magnetic field $\sim 10^{-16}$
  G over a comoving correlation length $\sim 10^{-2}$ pc is
  predicted,  our model predicts a present magnetic field $\sim 10^{-7}$ G
  over a comoving length $ \sim$ 1 pc. The predicted magnetic field
  in our model is , thus, nine orders of magnitude greater (over a
  comoving length two orders of magnitude greater) than that in the
  turbulent magnetogenesis model. This large difference is due to
  the fast energy transfer from large to small dissipation scales in
  a Hubble time in the turbulent magnetogenesis model, which does
  not occur in our model.

\section{Creation of Magnetic Fields Due to the Electromagnetic Fluctuations in Hot Dense Equilibrium Primordial Plasmas}

Thermal electromagnetic fluctuations are present in all plasmas,
including those in thermal equilibrium, the level of which is
related to the dissipative characteristics of the medium, as
described by the Fluctuation-Dissipation Theorem (FDT) \citep{kub57}
[see also \citet{akh75,sit67,Ros65,daw68}]. The spectrum of the
fluctuations of the electric field  is given by
\begin{equation}
\frac{1}{8} \langle E_{i}E_{j} \rangle_{k\omega}=
\frac{i}{2}\frac{\hbar}{e^{\hbar\omega/T-1}}(\Lambda_{ij}^{-1}-\Lambda_{ij}^{-1*}),
\end{equation}

\begin{equation}
\Lambda_{ij}(\omega,\textbf{k})=\frac{k^{2}c^{2}}{\omega^{2}}\left(\frac{k_{i}k_{j}}{k^{2}}-\delta_{ij}\right)+\varepsilon_{ij}(\omega,\textbf{k}),
\end{equation}
where $\varepsilon_{ij}(\omega,\textbf{k})$ is the dielectric tensor
of the plasma, $\omega$  the frequency, and $\bar{\textbf{k}}$ is
the wave number of the fluctuation. From Faraday's law,
$\textbf{B}$ = $c\textbf{k}/\omega\times\textbf{E}$, and setting
$\textbf{k}$ = $k\hat{\textbf{x}}$, we find for the perpendicular
$B_{2}$ and $B_{3}$ magnetic fluctuations:
\begin{equation}
\frac{\langle B_{2}^{2} \rangle_{k\omega}}{8\pi} =
\frac{i}{2}\frac{\hbar}{e^{\hbar\omega/T-1}}\frac{c^{2}k^{2}}{\omega^{2}}(\Lambda_{33}^{-1}-\Lambda_{33}^{-1*}),
\end{equation}
and
\begin{equation}
\frac{\langle B_{3}^{2} \rangle_{k\omega}}{8\pi} =
\frac{i}{2}\frac{\hbar}{e^{\hbar\omega/T-1}}\frac{c^{2}k^{2}}{\omega^{2}}(\Lambda_{22}^{-1}-\Lambda_{22}^{-1*}),
\end{equation}
where the subscripts 1, 2, and 3 refer to the x, y, z directions. We
then have for the total magnetic fluctuations:
\begin{equation}
\frac{\langle B^{2} \rangle_{k\omega}}{8\pi} =
\frac{i}{2}\frac{\hbar}{e^{\hbar\omega/T-1}}\frac{c^{2}k^{2}}{\omega^{2}}(\Lambda_{22}^{-1}+\Lambda_{33}^{-1}-\Lambda_{22}^{-1*}-\Lambda_{33}^{-1*}).
\end{equation}
In order to obtain $\Lambda_{ij}(\omega,\textbf{k})$ from the
equations of motion of the plasma, a multifluid model for the plasma
is introduced,
\begin{equation}
m_{\alpha}\frac{d\textbf{v}_{\alpha}}{dt} =
e_{\alpha}\textbf{E}-\eta_{\alpha}m_{\alpha}\textbf{v}_{\alpha},
\end{equation}
where $\alpha$ is a particle species label and $\eta_{\alpha}$, the
collision frequency of the species.  From a Fourier transformation
of the above equation and rearranging terms, the dielectric tensor
can be obtained:
\begin{equation}
\epsilon_{ij}(\omega,\textbf{k}) =
\delta_{ij}-\sum_{\alpha}\frac{\omega_{p\alpha}^{2}}{\omega(\omega+i\eta_{\alpha})}\delta_{ij},
\label{permissividade}
\end{equation}
 where $\omega_{p\alpha}$ is the plasma frequency of the species $\alpha$. For, an
electron-positron plasma, the plasma frequency of the electrons is
equal to that of the positrons, $\omega_{pe^{+}} = \omega_{pe^{-}}$,
and the collision frequencies of the electrons and positrons are
equal, $\eta_{e^{+}} = \eta_{e^{-}} = \eta$. The dielectric tensor
from Eq. (\ref{permissividade}) then becomes
\begin{equation}
\epsilon_{ij}(\omega,\textbf{k}) =
\delta_{ij}-\frac{\omega_{p}^{2}}{\omega(\omega+i\eta)}\delta_{ij},
\end{equation}
where $\omega_{p}^{2} = \omega_{pe^{+}}^{2}+\omega_{pe^{-}}^{2}$.
For electrons, the Coulomb collision frequency is $\eta_{e} =
2.91\times10^{-6}n_{e}\ln\Lambda T^{-3/2}$ $(eV)s^{-1}$, where
$n_{e}$ is the electron  density.  The collision frequency for the
case of an electron-proton plasma, which dominates after the
primordial nucleosynthesis era, is $\eta_{p} =
4.78\times10^{-18}n_{e}\ln\Lambda T^{-3/2}$ $(eV)s^{-1}$. It
describes the binary collisions in a plasma, which we assume to be
the dominant contribution to $\eta$. We then obtain
\begin{equation}
\Lambda_{ij}= \left(\begin{array}{ccc}
  1-\frac{\omega_{p}^{2}}{\omega(\omega+i\eta)} & 0 & 0 \\
  0 & 1-\frac{c^{2}k^{2}}{\omega^{2}}- \frac{\omega_{p}^{2}}{\omega(\omega+i\eta)} & 0 \\
  0 & 0 & 1-\frac{c^{2}k^{2}}{\omega^{2}}-
  \frac{\omega_{p}^{2}}{\omega(\omega+i\eta)}
\end{array}\right).
\end{equation}
From Eqs.(7)-(11), the total magnetic field fluctuations as a
function of frequency and wave number k were found to be
\begin{equation}
\frac{\langle B^{2} \rangle_{k,\omega}}{8\pi} =
\frac{2\hbar\omega}{e^{\hbar\omega/T}-1}\eta\omega_{p}^{2}\frac{k^{2}c^{2}}{(\omega^{2}+\eta^{2})k^{4}c^{4}+2\omega^{2}(\omega_{p}^{2}-\omega^{2}-\eta^{2})k^{2}c^{2}+[(\omega^{2}-\omega_{p}^{2})^{2}+\eta^{2}\omega^{2}]\omega},
\label{bfinal}
\end{equation}
\citep{taj92}.

\section{Our Model}

Our model  is based on the magnetic fluctuations in the plasma
created  immediately after the QHPT, which are  described by the FDT
in the previous section. This plasma was composed primarily of
electrons, photons, neutrinos, muons, baryons and their
antiparticles. The baryons were essentially stationary and did not
contribute  to the fluctuations while the muons also contributed
very little and
 for a very short time. Since neutrinos are essentially massless
 and act qualitatively like photons, albeit  with much smaller
 cross sections, we assume that they also affect the magnetic fluctuations  very little.
Therefore we consider only an electron-positron-photon plasma before
the electron-positron annihilation era and an electron-proton plasma
thereafter.

 Most of the electromagnetic fluctuations in the primordial plasma that were created immediately after the
 QHPT fall into two
broad categories: those with large wavelengths (k $\lesssim$
$\omega_{pe}$/c) at near zero frequency ($\omega$ $\ll$
$\omega_{pe}$) and those with very small wavelengths (k $\gg$
$\omega_{pe}$/c) and frequencies greater than $\omega_{pe}$. The
modes $k \lesssim \omega_{pe}/c$, denominated  ``soft" or ``plastic"
photons by \citet{taj92}, were significantly modified.  It is these
plastic photons and their magnetic fields in which we are
interested.

 From Eq.(\ref{bfinal}), we obtain the strength
of the magnetic field whose wavelengths are larger than a
 size $\lambda$,

\begin{equation}
\langle B^{2} \rangle_{\lambda}/8\pi = (T/2)(4\pi/3)\lambda^{-3},
\label{blambda}
\end{equation}
which decreases rapidly with wavelength. Thus, the magnetic field in
Eq.(13) was concentrated near the wavelength $\lambda$. The spatial
size $\lambda$ of the magnetic field fluctuations is related to
$\tau$, the lifetime of the fluctuation, by
\begin{equation}
\lambda(\tau) = 2\pi\frac{c}{\omega_{p}}(\eta_{e}\tau)^{1/2}
\end{equation}
\citep{taj92}. The average size of the magnetic fluctuations was
 \begin{equation}
 \bar{\lambda} = \frac{\int\lambda[\langle B^{2} \rangle_{\lambda}/8\pi]d \lambda}{\int [\langle B^{2}
 \rangle_{\lambda}/8\pi]d \lambda} = \frac{7\pi}{3}(c/\omega_{p}).
 \label{mediumsize}
 \end{equation}

 Using the model of \citet{taj92}, we assume that a fluctuation
 predicted by the FDT can be
 described by a bubble of size $\bar{\lambda}$. It contains a
 magnetic dipole whose field intensity is given by Eq.
 (\ref{blambda}).

The magnetic bubbles were at the temperature of the plasma. We
assume that  they touched each other and coalesced in a time
$t_{coal} = \bar{\lambda}/v_{bub}$, where $v_{bub}$ was the thermal
velocity of the bubble. The coalescence time $t_{coal}$ was found to
be much shorter than the lifetime $\tau$ of the bubbles in the
primordial Universe, for example, $\sim 10^{-5} $s shortly after t $
\sim 10^{-4}$ s. Before the magnetic fields dissipated, the bubbles
coalesced with one another. Once a larger
 bubble was formed, its lifetime, which  is
proportional to the square of its size, was longer. Larger bubbles
lived longer and, thus, had more  opportunities to collide with
other bubbles. In this way, a preferential formation of larger
bubbles occurred.

 Magnetic field fluctuations are created immediately after the QHPT
as predicted by the FDT, which we evolve to the recombination era
and beyond. Magnetic field fluctuations are also predicted to be
created by the FDT at the recombination era. Since the created
magnetic field fluctuations $\langle B \rangle^{2}$are proportional
to $Tn^{3/2}$,  the evolved magnetic fields from the QHPT at the
recombination era are very much greater than the created magnetic
fields at the recombination era. The latter source of primordial
magnetic fields was thus neglected in our investigation.
\citet{taj92} previously suggested that the evolved primordial
fields might continue to exist at the present epoch. No explicit
calculation was, however, made. Thus, previously it was not known
whether these fields would continue to exist or not to the present
era. We show here that these fields do indeed continue to exit and
are not destroyed in their evolution by diffusion or reconnection.
We also evaluate  their structure and intensity as a function of
redshift.

 We begin our calculations immediately after the QHPT and continue to  $\emph{z} \sim
 10$. Magnetic fields were adiabatically amplified at \emph{z} $\sim$ 10 as the baryon matter
 collapsed to form galaxies.

 Although the magnetic  energy density of neighboring magnetic dipoles
is of the same order as the energy density of the average magnetic
field when  they are not at the average distance from each other,
the magnetic energy density appreciably increases when the
neighboring dipoles approach each other. Since the field of a dipole
is proportional to $r^{-3}$, where r is the distance from the
dipole, the magnetic energy density of neighboring dipoles is
proportional to $r^{-6}$. Decreasing the separation distance by a
factor of 2(4) for example, increases the energy density by a factor
64(4096). Thus the magnetic energy density of adjacent magnetic
bubbles at very short separation distances is very much greater than
the average magnetic energy density.

 The dipoles  tended to align as
 they
  interacted due to the intensification of the magnetic interaction energy at shorter inter-dipole-distances.
  The interaction rate of the dipoles
  depended
   on  their thermal velocity. We used as the thermal velocity
the  velocity of the mass of the plasma bubble which is in thermal
equilibrium at the temperature of the Universe at a given redshift.
When
 the dipoles were oppositely oriented and interacting , two opposing processes
 occurred:  alignment and reconnection. As the dipoles approached each
 other, they tended to align in a flip time $\tau_{flip} \sim 10^{-5}$
 s shortly after the QHPT
 at t $ \sim 10^{-4}$ s, where $\tau_{flip}$ is the time
 in which a bubble  aligns with a neighboring bubble due to the magnetic
 torque. We have $\tau_{flip} \propto \left(I/N_{mag}\right)^{1/2}$, where
$N_{mag}$ is
 the magnetic torque and I is  moment of inertia  of the bubble. On the other
 hand, the opposite magnetic fields of the dipoles  reconnected
 in a tearing mode time $\tau_{tear}$. The shortest
 $\tau_{tear}$ is estimated to be  $\bar{\tau}_{tear}\cong
 10^{0.20}\tau_{A}^{1/2}\tau_{R}^{1/2}$, where $\tau_{A} =
 L/v_{A}$ is the Alfv\'{e}n time and $\tau_{R} = 4\pi L^{2}/c\eta$
 is the resistive time \citep{stu94}.  The shortest tearing time shortly after the QHPT was $\sim 10^{15}$ s. Thus
 $\tau_{flip}\ll\bar{\tau}_{tear}$ shortly after $10^{-4}$ s and remains so for all times of interest.
  Fig. (1) plots $\tau_{flip} $, Fig. (2) plots
  $\tau_{tear} $ and Fig. (3) their ratio,
  in the time interval $\sim 10^{-4}-10^{2}$ sec.

 The final time plotted in Figs. (1)-(3), $\sim 100$ s, is  the time in which
   the magnetic field in a bubble requires the age of the
 Universe to diffuse away. Magnetic diffusion, inversely
 proportional to the square of the diameter of the bubble, is only
 important at early times, when the bubbles were small. An initial
 magnetic field in a  bubble diffused away  in a time $\tau_{diff} = 4\pi\sigma L^{2}$, where L is
 the diameter of the bubble and $\sigma$ is the electrical
 conductivity \citep{Gra01}.

 In the high temperature regime (T $>$ 1 MeV) we
 followed the treatment of \citet{aho96} who solved numerically the Boltzmann equation in
 the early Universe.   For $T \lesssim 100$ MeV they found for the conductivity $\sigma \simeq
 0.76T$.  Since immediately after the QHPT
  the temperature of the Universe
  was $\sim 100$ MeV, we used $\sigma \simeq 0.76T$ for T $>$ 1 MeV.

  At temperatures T $<$ 1 MeV the conductivity can be approximated as

  \begin{equation}
  \sigma = \frac{m_{e}}{\alpha \ln\Lambda}\left(\frac{2T}{\pi
  m_{e}}\right)^{3/2},
  \end{equation}
where $\Lambda =
(1/6\pi^{1/2})(1/\alpha^{1/2})(m_{e}^{3}/n_{e})^{1/2}(T/m_{e})$, and
$\alpha, m_{e}$, and $n_{e}$ is the fine structure constant, the
electron mass, and the electron density, respectively \citep{jac75}.
For L $\sim$ 1 A.U., $\tau_{diff}$ is
 equal to  the age of the Universe \citep{Gra01}. In our
 model the bubbles reached a size $\sim$ 1 A.U. in a time $\sim$ 100
 s. In Figs. (1)-(3)  $\tau_{flip}$ and $\tau_{tear}$ are
  thus plotted  from the time of the QHPT ($\sim 10^{-4}
 s$) to $\sim$ 100 s.

 The magnetic field in a bubble would dissipate before coalescence of the bubble occurred
 if the magnetic diffusion time was smaller than the
 coalescence time.  In Fig. (4) we plot the ratio of the coalescence time $\tau_{coal}$ to the
 diffusion time $\tau_{diff}$. It can be seen in Fig. (4) that this ratio remains very much less than unity at early
 times.

  At late times, when the magnetic field flip time (i.e., the time for adjacent dipoles
to align) was greater  than the Hubble time, the magnetic dipoles
remained random. The transition redshift, when random fields began
to exist, was $\emph{z} \sim 10^{8}$. At this epoch, the comoving
  size of the bubbles
 was $\sim$ 1 pc. In order to explain galactic magnetic fields, we need to evaluate
 the field over the comoving scale of a protogalaxy,
 $\sim$ 1Mpc, which eventually  collapsed to the comoving scale of a galaxy, $\sim 30$ kpc.

 The magnetic field in a bubble decreased adiabatically as the Universe expanded. Since magnetic
flux is
 conserved, we have
\begin{equation}
B = \frac{B_{0}}{a^{2}},
\end{equation}
where a is the cosmic scale factor. A $\Lambda$CDM model was used to
evolve $a$ as function of time,  with $\Omega_{m}$ = 0.3,
$\Omega_{\Lambda}$ = 0.7, and a Hubble constant $h\equiv$ H/100 km
$s^{-1} Mpc^{-1}$ = 0.72.

   In Figs. (5) and (6), we show the evolution of the size of the  bubbles  as a function of
   time, from immediately after the QHPT at $10^{-4}$ s to a
   redshift $\emph{z} \sim 10$ at a time $\sim 10^{16}$ sec. Initially, the size of the bubbles increased rapidly,
   as shown in Fig. (5).
    From Fig.(5), we observe
   that the physical size of a bubble increased from $10^{-8}$ cm
   at
   $t \approx 10^{-4}$ to 1 cm in a time $10^{-7}$ sec. It
    continued to increase at this rate until it reached a size
   $\sim 10^{7}$ cm. The growth rate then decreased, as shown in
   Fig. (6). At the redshift $\emph{z} \sim 10^8$ ($t \sim 3000 s$),
   the physical size of the bubble was $\sim 10^{10}$ cm (i.e., a comoving size $\sim$ 1
   pc).

The manner in which we extrapolated the field amplitude to
cosmological scales
  followed the phenomenological analysis
 of random distributions of size L, in the review article of \citet{Gra01}.
 Their generic average magnetic field over a distance D at a time t is proportional to
 $(L/D)^{p}$, where p = 1/2, 1 or 3/2,

\begin{equation} \langle B(L,t) \rangle_{rms} = B_{0}
\left(\frac{a_{0}}{a(t)}\right)^{2} \left(\frac{L}{D}\right)^{p}.
 \label{medium}
\end{equation}
If we are interested in the volume average magnetic field of a
random distribution of dipoles in a sphere  of diameter D, and each
dipole is in a cell of diameter $L$, the average magnetic field is
proportional to $(L/D)^{3/2}$ and  p = 3/2 in Eq.(\ref{medium})
\citep{wid02}. If, however, we are interested in a line-of-sight
average magnetic field felt by a cosmic ray particle or a photon
(e.g. in Faraday Rotation Measurements) the average magnetic field
is proportional to $(L/D)^{1/2}$, and p = 1/2  in Eq.(\ref{medium}).

 Non-negligible volume average magnetic field can be important in
the formation of the first stars and in reionizing the Universe. The
formation of the first objects marks the transformation of the
Universe from its smooth initial state to its clumpy current state.
In popular cosmological models, the first sources of light began to
form at a redshift \emph{z} $\sim$ 30 and reionized most of the
hydrogen in the Universe by \emph{z} $\sim$ 7 \citep{bar01}. In
general, it is found difficult to reionize the Universe with a
standard Salpeter initial mass function for the first stellar
sources formed by a standard fluctuation dark matter spectrum
\citep[]{cen03,fuk03,cia03,som03,hai03}. Primordial magnetic fields
produce additional fluctuations of baryons by the Lorentz force
\citep{tas06}. The magnetic tension is more effective on small
scales where the entanglements of magnetic fields are larger.
\citet{tas06} found that ionizing photons from Population III stars
formed in dark halos could easily have reionized the Universe by
\emph{z} $\simeq 10-20$ if the present intensity of the primordial
magnetic field is $B_{0} \sim$ nG on a comoving scale $\sim 0.1$
Mpc. The relevant Lorentz force causing the collapse of baryon
matter is proportional to

\begin{equation}
\vec{\nabla}\cdot\left[(\vec{\nabla}\times \vec{B}_{0}(\vec{x})
)\times \vec{B}_{0}(\vec{x})\right] \sim
\frac{B_{0}^{2}}{D^{2}}\equiv F.
\end{equation}

Thus \citet{tas06} found that a value $F \sim 10^{-28}G^{2}/pc^2$ is
important in forming the first objects. In our model a present
volume average magnetic field over a comoving scale $D$ is $B_{0}
\sim 0.1 \mu$G $(D(pc))^{-3/2}$. We thus have
$F=[10^{-14}/D^{2}]{G^{2}/pc^{2}}$ and obtain a \citet{tas06} value
of F  with D $\sim$ kpc.  We thus find that a  $D \sim$ kpc comoving
region in our model produces a Lorentz force which could be
important in forming the first stellar sources and in reionizing the
Universe. This length is larger than the magnetic Jeans length and
the cut off length due to direct cascade. Their respective wave
numbers, given by \citet{tas06}, are $k_{MJ} \sim 32 Mpc^{-1}
B_{0}^{-1} (nG).$ and $k_{c} \sim 102 Mpc^{-1} B_{0}^{-1} (nG)$.
Putting our volume average magnetic field $B_{0}$ over $D \sim 1$
kpc into these expressions we obtain $k_{MJ} \sim 10 kpc^{-1}$ and
$k_{c} \sim 34 kpc^{-1}$.  It is to be noted that a sphere of
comoving diameter $\sim 1$ kpc contains a mass $\sim 10^{3}
M_{\odot}$ for a reduced matter density $\Omega_{m} \sim 0.3$ and
Hubble parameter h $\sim$ 0.72.

 A detailed discussion on average procedures of tangled magnetic fields
can be found in \citet{hin98}. Table (1) shows the growth of the
magnetic field in our model and the size of the bubbles down to the
redshift $\emph{z}\sim 10$. The equipartition redshift  in Table (1)
was obtained from the relation $(1 + \emph{z}_{eq})\approx 2.3
\times 10^{4} \Omega_{m}h^{2}$ \citep{pad93}. Table (2) shows the
growth of the line-of-sight average magnetic field over a comoving
protogalactic size L $\sim 1$ Mpc.

 At \emph{z} = 10, the intensity of the magnetic field in a bubble
whose comoving size is $\sim 1$ pc was $\sim 9 \mu$G. Taking the
line-of-site average  over the comoving
 scale of 1 Mpc  ($\sim$ 100 kpc at \emph{z}
$\sim$ 10), the rms magnetic field at \emph{z} = 10 was   $9 \times
10^{-3}$ $\mu$G. The magnetic field in the bubbles as a function of
time is shown in Fig. (7). In Fig. (8), the evolution of the
line-of-sight average and volume average magnetic field of comoving
size $\sim$ 1 Mpc is shown as a function of time from $t\simeq 3
\times 10^{3}$ sec, when random fields began to exist, to \emph{z}
$\sim$ 10. In the collapse of the comoving 1 Mpc region at \emph{z}
= 10 to a galaxy (comoving size $\sim 30$ kpc), the field is
amplified to $\sim 10 \mu$G. This indicates that the magnetic fields
created immediately after the QHPT could be the origin of the $\sim
\mu$G fields observed in galaxies at high and low redshifts.

\section{Conclusions and Discussion}

We showed that the electromagnetic fluctuations in the primordial
plasma immediately after QHPT constitute a strong candidate for the
origin of primordial magnetic fields in galaxies and clusters of
galaxies. We calculated the magnetic field fluctuations in the
plasma after this transition  and evaluated their evolution with
time. Intense magnetic field fluctuations on the order of $10^{16}$
G existed at $t = 10^{-4}$ sec after the QHPT. These fields formed a
spatial linkage due to the process of successive coalescence. We
showed that magnetic bubbles created immediately after the
transition could survive to $\emph{z}\sim 10$ and could explain the
observed magnetic fields at high and  low redshifts determined by
Faraday Rotation Measurements (FRMs). We found: 1) $\sim 10 \mu$G
magnetic fields (MFs) over a comoving $\sim$ 1 pc region at a
redshift \emph{z} $\sim$ 10; 2) Line-of-sight average MFs, important
in Faraday Rotation Measurements, $\simeq 10^{-2} \mu$G over a 1 Mpc
comoving region at \emph{z} $\sim$ 10 which in the collapse to a
galaxy comoving size $\sim 30$ kpc, are amplified to $\sim 10 \mu$G;
and 3) Volume average MFs over a comoving 1 kpc region that
 could be important in forming he first stellar sources
  and in reioninzing the Universe \citep[]{sub06,tas06}.

We found that the magnetic fields in the  bubbles, created
originally at the QHPT, had a value $\sim 10 \mu$G at the redshift
$\emph{z} \sim 10$ and  a size 0.1 pc (Table 1). At the present
time, these bubbles have a comoving length $\sim$ 1 pc and a field
$\sim 0.1 \mu$G. We can compare these results with previous
calculations of the creation of magnetic fields at the QHPT.
\citet{che94}, for example, found a much smaller magnetic field,
$\sim 10^{-10} \mu$G, over the same comoving size with their
mechanism.
 \citet{qua89} also found
a much smaller resultant magnetic field, $\sim 2 \times 10^{-11}
\mu$G, over a much smaller comoving size, $\sim 10^{-5}$ pc.

It is to be noted that the origin of primordial magnetic fields
suggested here is qualitatively different from  the other previous
suggestions discussed in \S 3.1-\S 3.7. These previous suggestions
require special physical initial conditions. Our model, however,
does not. The magnetic fields in our model arise from the natural
fluctuations in the equilibrium plasma that existed in the
primordial Universe, described by the
Fluctuation-Dissipation-Theorem.

In \S 3.8, we discussed the model of \citet{tak05} which, like our
model, is based on natural fluctuations that exist in nature. Our
model, however, predicts very much larger magnetic amplitudes on a
comoving protogalactic scale $\sim$ 1 Mpc. \citet{tak05} found a
magnetic field B $\sim 10^{-25}$ G on a $\lambda$ = 10 Mpc comoving
scale. Since their field is $\propto k^{3}P(k)$, where k is the
wavenumber $(k = 2\pi/\lambda)$, and the fluctuation power spectrum
$P(k) \propto k^{n}$ with $n \sim -2$ for $\lambda < 10$ Mpc, the
\citet{tak05} prediction is $B \sim 10^{-23}$G at present for
$\lambda \sim$ 1 Mpc. This can be compared with our prediction for
the same comoving scale, which is
 many orders of magnitude greater.

 Our predicted magnetic fields are consistent with present observations.
Extragalactic magnetic fields as strong as $\sim 1 \mu$G in sheets
 and filaments in the large scale galaxy distribution, such as in
 the Local Superclusters, are compatible with existing FRMs
 \citep{sig03,kro94,val97,cla01,han02,ryu98,bla99}. These limits are
 consistent with our predicted fields. There is
mounting evidence from diffuse radio-synchrotron clusters
\citep{gio00} and  a few cases of filaments \citep{kim89,bag02} that
magnetic fields $\sim 0.1 - 1.0 \mu$G exist in the low density
outskirts of cosmological collapsed objects. These fields  may have
their origin in the primordial magnetic fields that we predict.

 In contrast to  the previous models suggested in \S 3, our model
 predicts relatively intense magnetic fields  over
 small regions  in the intergalactic medium. This
 prediction may help to solve the  long standing problem of
 ultra-high energy cosmic rays (UHECRs) ($> 3\times 10^{18}$ eV): their
 extreme isotropy. The UHECRs are extragalactic since their Larmor
 radii are comparable or greater than  the size of the Galaxy \citep{coc56,mor57}.
 There are only a few nearby sources  that could be the origin of these cosmic
 rays.
 However, the observed arrival directions of the UHECRs are highly isotropic \citep{aba04,wes04}.

 The importance of intergalactic magnetic fields in creating an
 isotropic distribution of UHECRs has been discussed in the
 literature. However, different articles arrived at opposite
 conclusions.
 Whereas \citet{far00} argue that the magnetic fields
 created the observed isotropic distribution, \citet{dol04,dol05}
 argue that they are unimportant. \citet{med01} argue that only weak
 intergalactic magnetic fields making
 small angular deflections of the UHECRs may
 be necessary, since the number of  UHECRs  sources may be much
 larger than  those that are presently observed and that it  is
 possible that fossil cocoons, so called radio ghosts, contribute to
 the isotropization of the UHECR arrival directions \citep{med01}.

 Primordial magnetic fields have been previously assumed to exist,
without an explanation for their origin. \citet{dol04,dol05}, for
example, assumed the existence of a homogeneous primordial magnetic
field $ \sim 10^{-3} \mu$G at \emph{z} $\sim 20$. They  made a
magnetohydrodynamic simulation of cosmic structure formation that
reproduces the positions of known galaxy clusters in the Local
Universe. Protons of energy $\geq 4 \times 10^{19}$ eV were found to
have deflections, which  do not exceed a few degrees over most of
the sky, up to a propagation distance of $\sim 500$ Mpc. It is
difficult to explain, however, an isotropic distribution of UHECRs
 with their analysis.

Relatively intense magnetic fields have been predicted to exist in
filaments in the intergalactic medium. Such a filament  might exist
between us and the powerful radio galaxy,  Cen A. For example
\citet{far00} suggested that  Cen A, at a distance of 3.4 Mpc, could
be the  source of most UHECRs observed. The extragalactic magnetic
field was estimated to be $\sim 0.3 \mu$G. They argue that this
scenario can account for the spectrum of UHECRs down to $\approx
10^{18.7}$ eV, including its isotropy and spectral smoothness.

If our predicted magnetic fields are not spread uniformly over space
but, as expected, are concentrated into the web of filaments,
predicted by numerical simulations, appreciable deflections of
UHECRs propagating along the filaments could occur. The deflection
in a distance D of a UHECR with energy $E \equiv E_{20} \times
10^{20}$ eV by
 magnetic bubbles of size $\lambda$ and magnetic field B  is
\begin{equation}
\delta(\theta) \sim
0.5^{0}\left[D(Mpc)\lambda(Mpc)\right]^{1/2}B(nG)/E_{20}
\end{equation}
\citep{far00}. We predicted magnetic bubbles with B $\sim 10 \mu$G
and $\lambda \sim 0.1$ pc at \emph{z} $\sim$ 10. Let us assume: that
$\lambda$ increased with the cosmic scale factor and that  $\lambda
\sim$ 1 pc  at \emph{z} $\sim$ 10; and that  the magnetic fields,
trapped in the filaments, decreased slightly to B $\sim 1-10 \mu$G.
From Eq. (18), with  a distance D $\sim$ 100 Mpc, we obtain
$\delta(\theta) \sim 5^{0}-50^{0}/E_{20}$. Appreciable deflections
could, thus, occur along filaments.

\begin{figure}
 \includegraphics{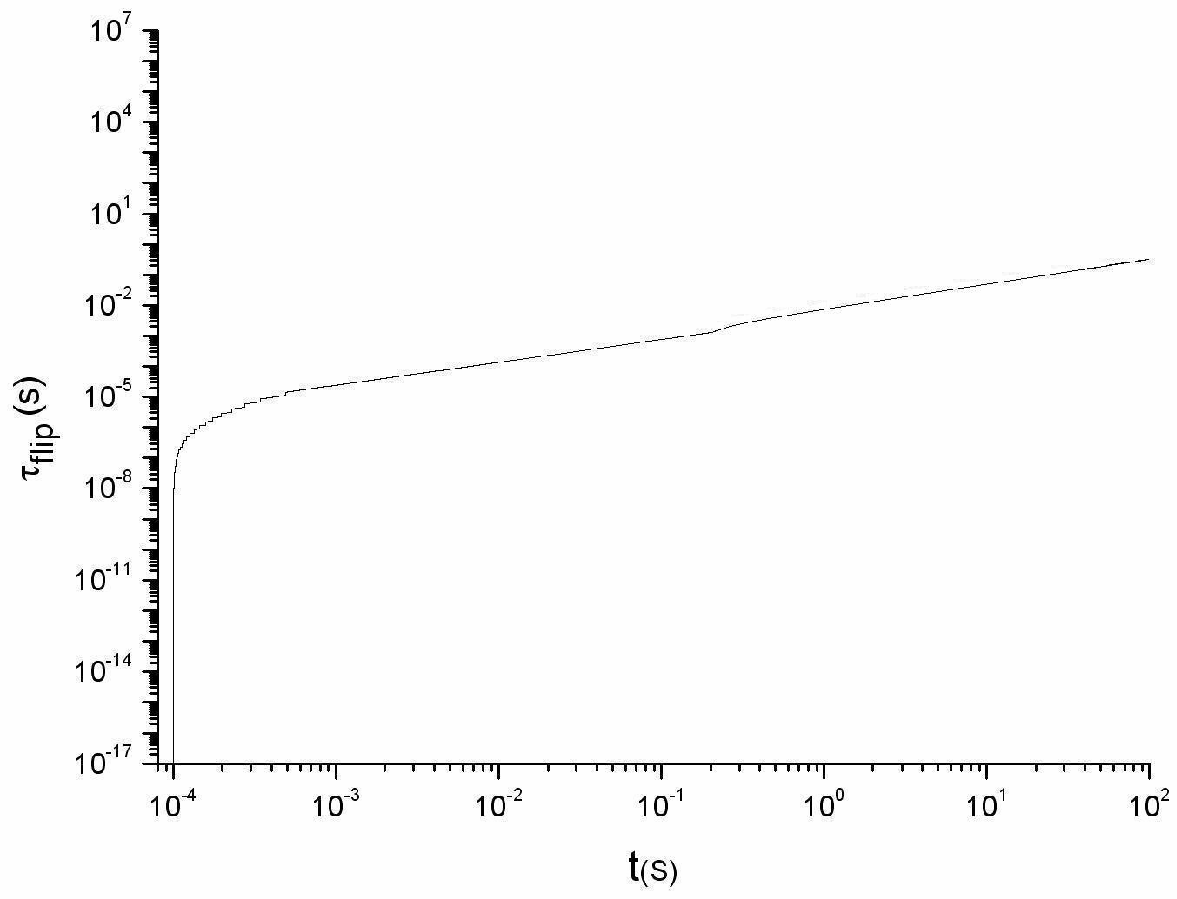}
 \caption{Evolution of the flip time $\tau_{flip}$ (s) of the bubbles  as a function of time t(s).}
\end{figure}
\clearpage

\begin{figure}
 \includegraphics{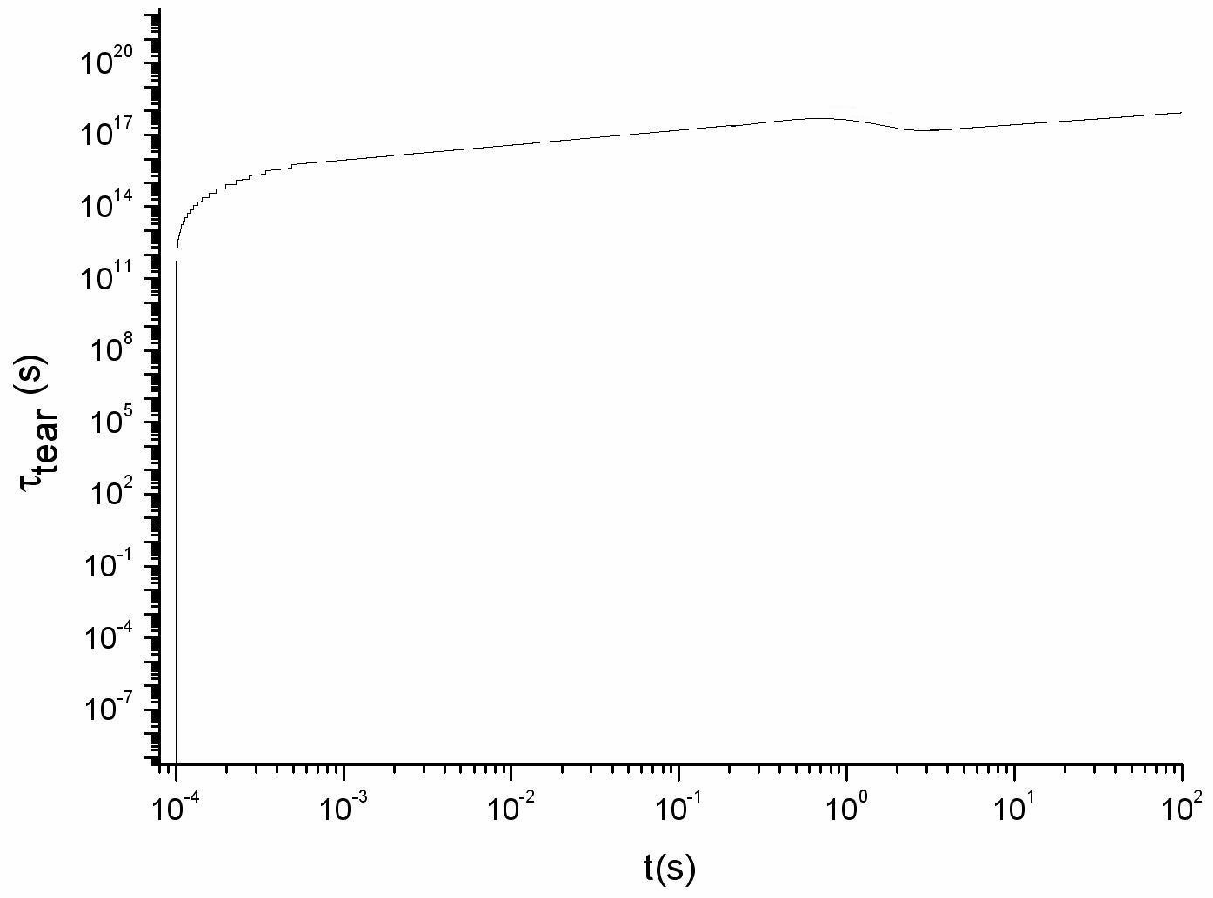}
 \caption{Evolution of the tearing time $\tau_{tear}$ (s) of the bubbles  as a function of time t(s).}
\end{figure}
\clearpage

\begin{figure}
 \includegraphics{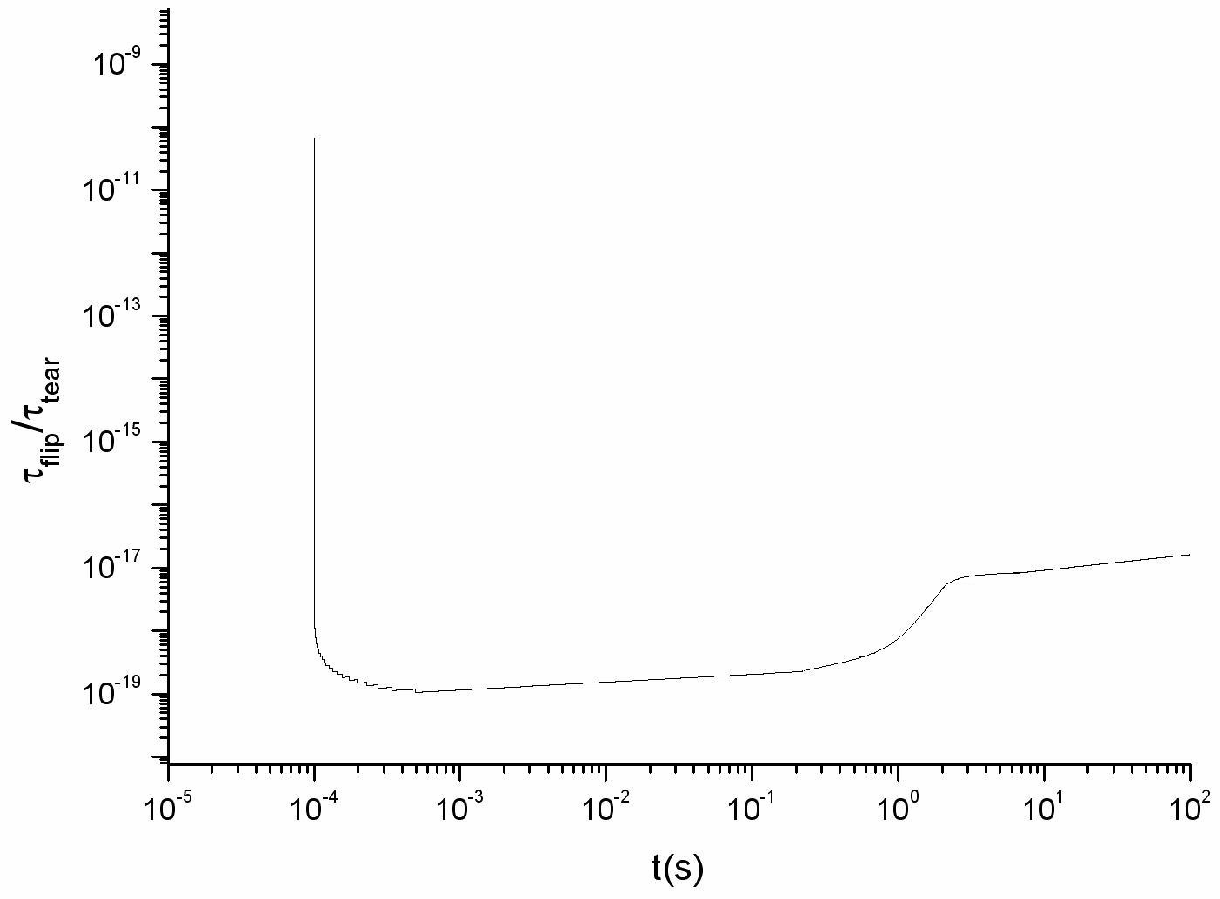}
 \caption{Ratio of the flip time $\tau_{flip}$  of the bubbles to the tearing time
 $\tau_{tear}$
  as a function of time t(s).}

\end{figure}
\clearpage

\begin{figure}
 \includegraphics{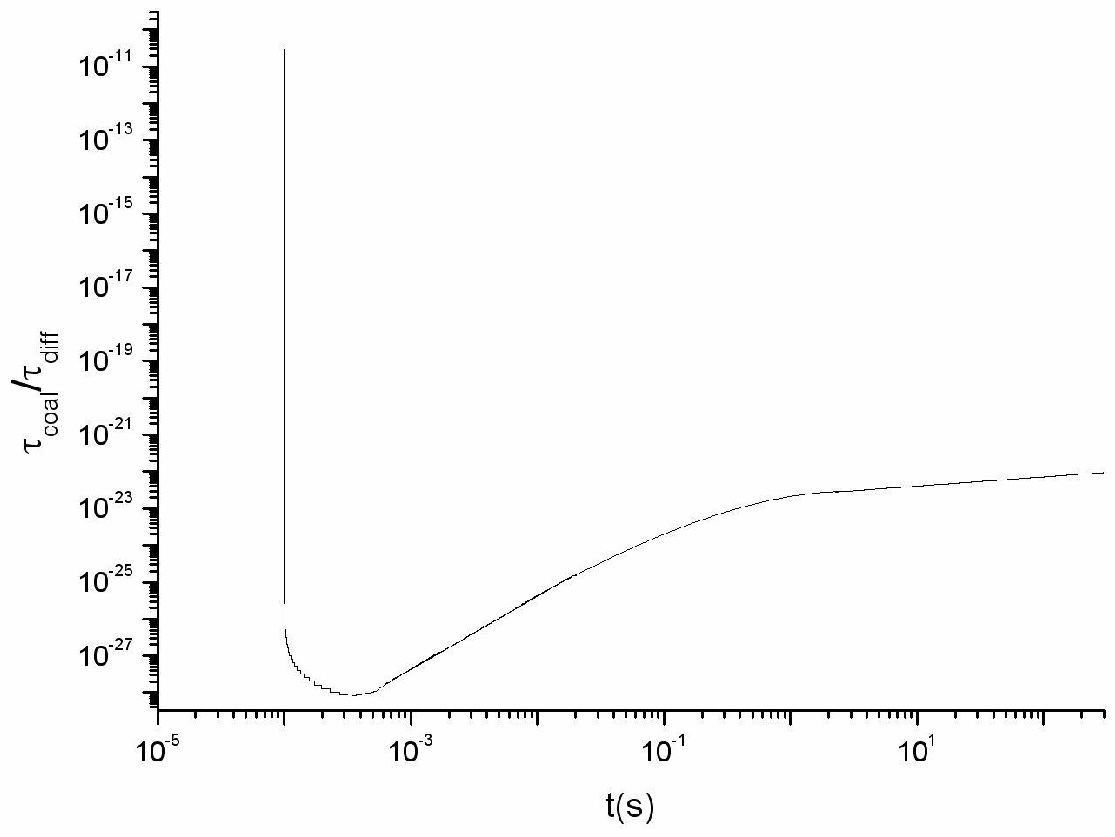}
 \caption{Ratio of the coalescence time $\tau_{coal}$  of the bubbles to the diffusion time
 $\tau_{diff}$
  as a function of time t(s).}

\end{figure}
\clearpage

\begin{figure}
 \includegraphics{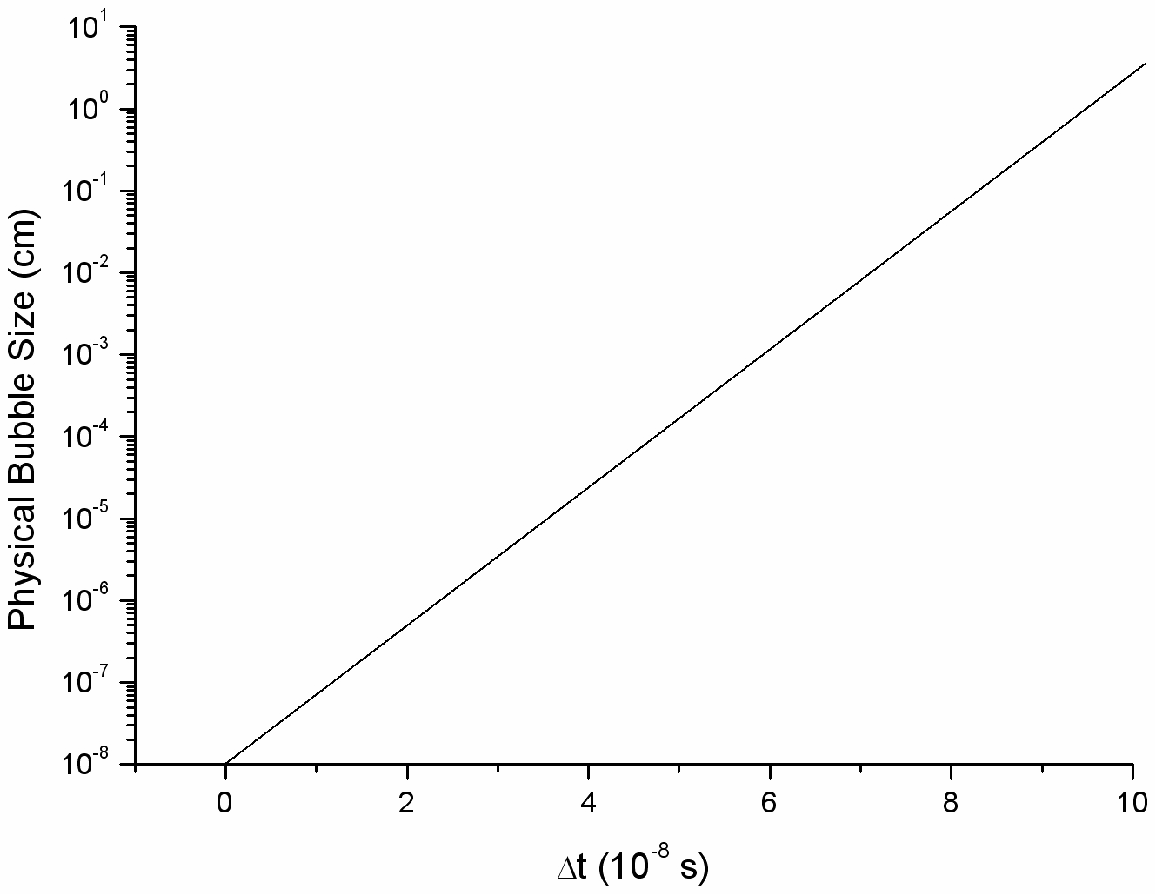}
 \centering
 \caption{Initial evolution of the physical size of the magnetic
 bubbles, created immediately after the QHPT,
 as a function of time, $t\equiv t_{0}+ \Delta t$, for  $t_{0} = 10^{-4}$
  s, and $0 < \Delta t (10^{-8} s) \leq 10$.}

\end{figure}
\clearpage

\begin{figure}
 \includegraphics{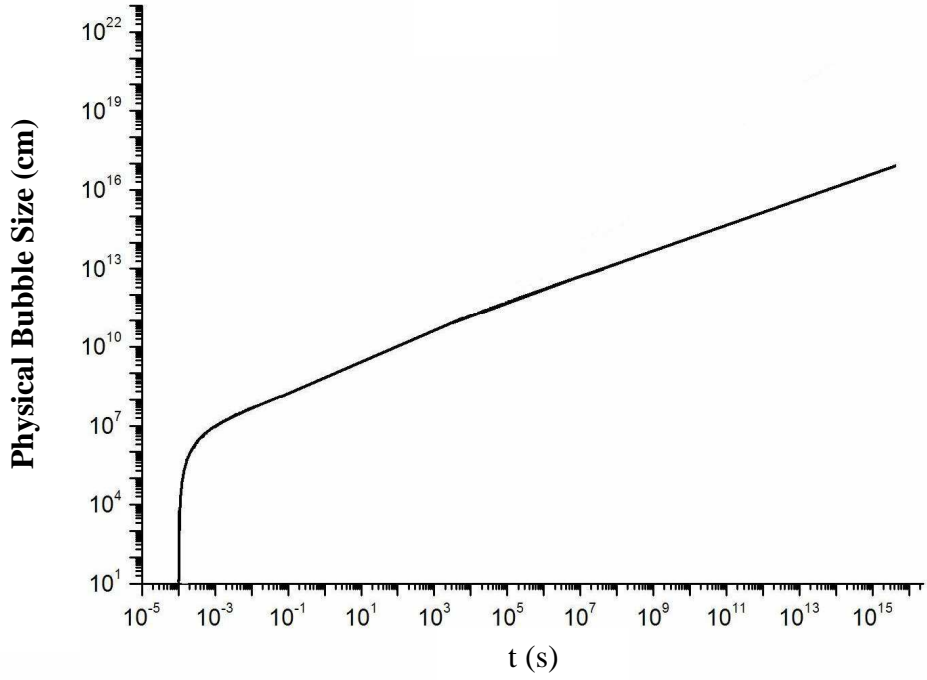}
 \caption{Evolution of the physical size of the magnetic bubbles as a function of time from t $\sim 0.1 s$.}

\end{figure}

\clearpage

\begin{figure}
 \includegraphics{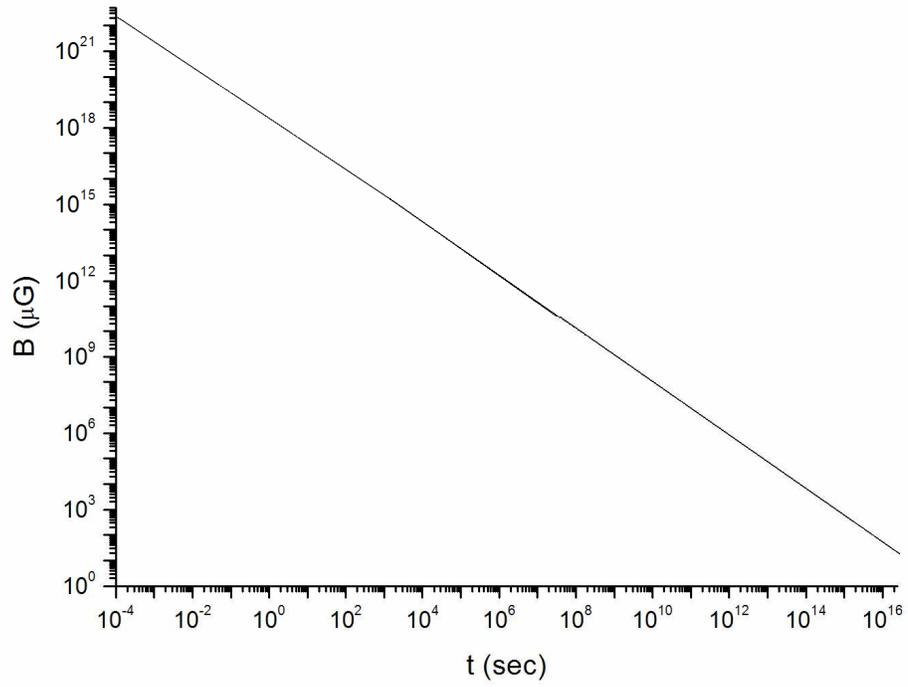}
 \caption{Evolution of the magnetic field $B(\mu$ G) in the bubbles, created immediately after the
 QHPT,
 as a function of time, t(sec).}
\end{figure}
\clearpage

\begin{figure}
 \includegraphics{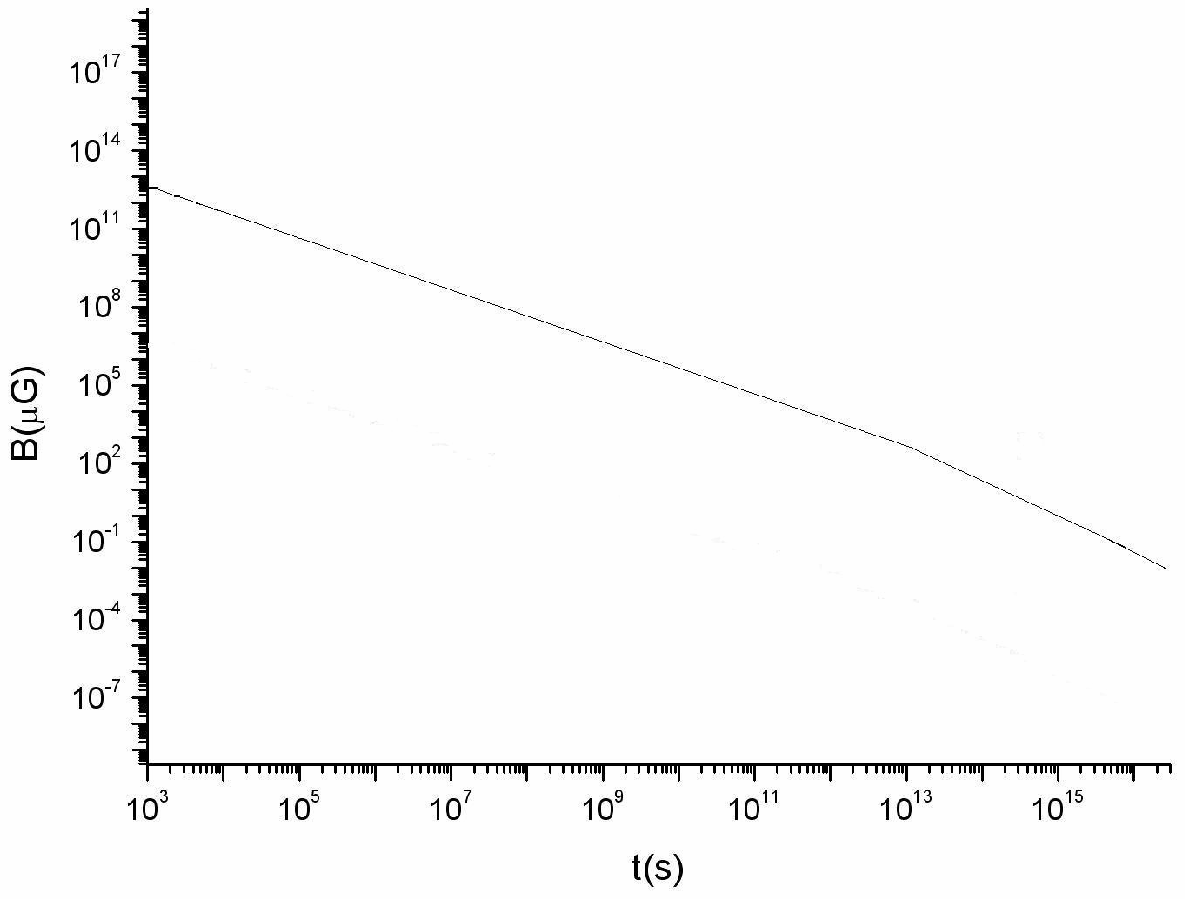}
 \caption{Evolution of the line-of-sight average  magnetic field $B(\mu$ G) of comoving size $\sim$ 1 Mpc  as a function of time t (sec) from $t\simeq 3 \times 10^{3}$ sec, when random
 fields began to exist, to   t $\sim 10^{16}$ sec (\emph{z} $\sim$ 10), when galaxies began to form.}

\end{figure}
\clearpage

\begin{table}
\caption{Size and Strength of Magnetic Fields in Bubbles}
\begin{ruledtabular}

\begin{tabular}{lcccc}
 Epoch & Magnetic Field ($\mu$G)   & Redshift   &
Time (sec)&Size (cm)\\
\hline

Immediately after the QHPT & $10^{22}$     & $6 \times 10^{11}$  &  $10^{-4}$ & $10^{-12}$\\
Electron positron annihilation era &   $10^{18}$    & $10^{10}$  & 1  & $10^{8}$ \\
Nucleosynthesis era &   $10^{15}$ &    $10^{8}-10^{9}$ & $1-500$
 & $10^{10}$  \\
Equipartition era & $2 \times 10^{5}$    & $3600$  & $10^{12}$
 & $3 \times 10^{14}$\\
Recombination era & $2 \times 10^{2}$  &  $1100$ & $8 \times
10^{12}$  & $10^{15}$ \\
Galaxy formation era &   9   &  $\sim 10$ & $10^{16}$ &
$10^{17}$  \\

\end{tabular}
\end{ruledtabular}
\end{table}

\clearpage

\begin{table}
 \caption{Line-of-sight Average Magnetic Fields in Protogalaxies of Comoving Size $\sim$ 1Mpc}
\begin{ruledtabular}
\begin{tabular}{lccccc}

Epoch & Magnetic Field ($\mu$G)   & Redshift    &
Time (sec) & Size (cm) \\
\hline

Beginning of random fields  &  $9.5 \times 10^{11}$  &   $10^{8}$  &  $3 \times 10^{3}$  & $10^{-12}$ \\
Equipartition era &   $10^{4}$ &  $3600$  & $10^{12}$  &
$10^{18}$  \\
Recombination era &  $300$  &   $1100$ & $8 \times 10^{12}$ &
$4 \times 10^{22}$ \\
Galaxy formation era &    $9 \times 10^{-3}$ &     $\sim 10$ &
$10^{16}$&
$10^{23}$ \\
\end{tabular}
\end{ruledtabular}
\end{table}
\clearpage

\acknowledgments

R.S.S. thanks the Brazilian agency FAPESP for financial support
(04/05961-0). R.O. thanks FAPESP (00/06770-2) and the Brazilian
agency CNPq (300414/82-0) for partial support.

 \clearpage

\end{document}